\documentstyle[12pt]{article}

\begin{document}
\vspace{-2.0cm}
\bigskip
\begin{center}
{\Large \bf
Wigner's Little  Group,  Gauge Transformations and Dimensional Descent} 

\end{center}
\vskip .8 true cm
\begin{center}
{\bf Rabin Banerjee}\footnote{rabin@boson.bose.res.in} and
{\bf Biswajit Chakraborty}\footnote{biswajit@boson.bose.res.in}

\vskip 1.0 true cm

S. N. Bose National Centre for Basic Sciences \\
JD Block, Sector III, Salt Lake City, Calcutta -700 098, India.
        
\end{center}
\bigskip

\centerline{\large \bf Abstract}
We propose a technique called dimensional descent  to show that Wigner's
 little group for massless particles, which acts as a generator of gauge 
transformation for usual Maxwell theory, has an identical role even 
for topologically massive gauge theories. The examples of $B\wedge F$ 
theory and Maxwell-Chern-Simons theory are analyzed in details.  
\vskip 1.0 true cm

Wigner's  little group\cite{w}, which is a subgroup of the Poincare group
that leaves the four-momentum invariant,
 is of fundamental importance for both massive and massless particles. In the latter case this group is $E(2)$, which is a semi-direct product of $T(2)$ (the   
group of translations in the plane) and $SO(2)$.
Each of these subgroups has its own role, but here we are concerned with $T(2)$
which acts as a generator of gauge transformation\cite{hk, we}  in  usual gauge theories,
like Maxwell or Kalb-Ramond\cite{kr}, where the quanta of excitations are massless.
There are however instances where gauge invariance coexists  with
mass - the topologically massive gauge theories like the $B\wedge F$\cite{cs}
theory or the Maxwell-Chern-Simons(MCS)\cite{djt}  theory in 2+1 dimensions.
Recently we had  
shown\cite{bc} that in the $B\wedge F$ theory, the gauge invariance is generated by a
particular representation of $T(3) \subset E(3)$, so that one has to go beyond
Wigner's little group. For the MCS model, however, Wigner's little group
(isomorphic to ${\cal R} \times {\cal Z}_2)$ itself suffices, albeit in a
different representation. Is there then a systematic method of obtaining
 these distinct
representations and, if possible, connect them to Wigner's little group 
for massless particles, which acts as gauge generator for conventional gauge
theories?

The object of this paper is to analyze this and related issues. We show that the specific representations of the translation groups that act as gauge 
generators of topologically massive gauge theories follow naturally by a
dimensional descent of the standard representation of Wigner's little group
in one higher dimension. As a byproduct we show the connection between the
helicity quantum numbers of  massless particles	 and the spin quantum number of
 massive particles in one lower dimension.  

Let us first recapitulate certain results\cite{bc} for the $B\wedge F$
theory in 3+1 dimensions, whose dynamics is governed by the following Lagrangian density\footnote{Greek indices will always denote 3+1 dimensional space-time.
 Latin indices from the middle of the alphabet(like $i, j, k$) will indicate 
4+1 dimensions while those from the beginning(like $a, b, c$) will correspond
to 2+1 dimensions.}\cite{cs}
\begin{equation}
{\cal L} = 
-{\frac{1}{4}}{F^{\mu \nu}F_{\mu \nu}} + \frac{1}{12}H_{\mu \nu \lambda}
H^{\mu \nu \lambda} - \frac{m}{6}\epsilon_{\mu \nu \lambda \rho}H^{\mu \nu \lambda}A^{\rho}
\label{bfl}
\end{equation}
where,
\begin{equation}
F_{\mu \nu} = \partial_{[\mu} A_{\nu ]}; \hskip 0.5cm 
H_{\mu \nu \lambda} = \partial_{[\mu} B_{\nu \lambda ]}; \hskip 0.5cm
 \mu = 0, 1, 2, 3. 
\label{bfl-1}
\end{equation}
The polarization vector(tensor)  associated with the one(two) form fields
$A_{\mu} (B_{\nu \lambda})$ are given by\cite{bc},
\begin{equation}
\varepsilon^{\mu} = -i\left( \begin{array}{c} 
0 \\
a_1 \\
a_2 \\
a_3
\end{array}\right); \hskip  1.0cm \varepsilon = \{\varepsilon^{\nu \lambda} \}
= \left(
\begin{array}{cccc}
0 & 0 & 0 & 0 \\
0 & 0 & a_3 & -a_2 \\
0 & -a_3 & 0 &  a_1 \\
0 & a_2 & -a_1 & 0   
\end{array}\right)
\label{bfl-2}
\end{equation}
corresponding to the momentum four-vector,
\begin{equation}
p^{\mu} = (m, 0, 0, 0)^T
\label{bfl-3}
\end{equation}
for a massive quanta at rest. Notice that the entries in (\ref{bfl-2}) satisfy
a duality relation connecting the space components of the two matrices.

It might be mentioned that identical expressions for polarization vector(tensor)
are obtained for Proca( massive Kalb-Ramond) theory.\footnote{Of course the entries in the 
different matrices are now independent and are not constrained by the duality relation
mentioned earlier which is a consequence of the coupling term in the $B\wedge F $ model.}
 This is not unexpected
if one recalls the equivalence of the $B\wedge F $ model to such theories\cite{at}.

To begin with the present analysis, note that $T(3)$, which acts as the generator of 
gauge transformation in the $B\wedge F $ model\cite{bc},  is an abelian invariant
 subgroup of $E(3)$.  The group $E(3)$ is  
 Wigner's little group for a massless particle in 5 dimensions.
This suggests that we shall have to consider 5(= 4+1) dimensional space-time.
Now an element of Wigner's little group in 5 dimensions can be written as 
\begin{equation}
W_5(p,q,r; \psi, \phi, \eta) = 
\left( \begin{array}{ccccc}
1+ \frac{p^2 + q^2 + r^2}{2} & p & q & r & -\frac{p^2 + q^2 + r^2}{2} \\
p & & & & -p \\
q & &R(\psi, \phi, \eta) & & -q \\
r & & & & -r  \\    
\frac{p^2 + q^2 + r^2}{2} &  p & q & r & 1 -\frac{p^2 + q^2 + r^2}{2}   
\end{array}\right)
\label{ex-1}
\end{equation}
where $p,q,r$ are  any real numbers, while $R(\psi, \phi, \eta) \in
SO(3)$,  with $(\psi, \phi, \eta)$ being a triplet of Euler angles.
 The above result can be derived by following the standard treatment\cite{sw}.
 The 
corresponding element of the translational group $T(3)$ can be trivially 
obtained by setting $R(\psi, \phi, \eta)$ to be the identity matrix and 
will be denoted by $W_5(p,q,r) = W_5(p,q,r; {\bf 0})$. 

Let us now consider free Maxwell theory in 5-dimensions, 
\begin{equation}
{\cal L} = -{\frac{1}{4}}{F^{ij}F_{ij}} ; \hskip 1.0cm i, j = 0, 1, 2, 3, 4
\label{5ml}
\end{equation} 
 A free photon will have three independent
transverse degrees of freedom. Correspondingly the polarization vector
$\varepsilon^{i}$ can be brought to the following form,
\begin{equation}
\varepsilon^{i} = (0, a_1, a_2, a_3 , 0)^T 
\label{ex-2}
\end{equation}
if the photon of energy $`\omega $' is taken to be propagating in the
 4-direction so that the energy momentum 5-vector takes the following form
\begin{equation}
p^{i} = (\omega, 0, 0, 0, \omega)^T 
\label{ex-3}
\end{equation}
Note that this automatically satisfies the `Lorentz gauge' 
$\varepsilon^{i}p_i = 0$.
If we now suppress the last rows of the column matrices $\varepsilon^{i}$ (\ref{ex-2}) and $p^{i}$ (\ref{ex-3}) we end up with the  polarization vector and 
energy-momentum 4-vector of a Proca quanta of mass $m = \omega$ in the rest frame in 3+1 dimensions. This is equivalent to applying
the projection operator given by the matrix 
\begin{equation}
{\cal P}= diag(1, 1, 1, 1, 0)  
\label{ex-4}
\end{equation}
and deleting the last rows which now have only null entries. This method
of getting results in any dimension by starting from the known results
in one dimension higher will be called dimensional descent.
 
Similarly, to reproduce the polarization tensor of massive Kalb-Ramond(KR) 
theory in 3+1
dimensions let us consider free massless KR model in 5-dimensions,
\begin{equation}
{\cal L} = {\frac{1}{12}}{H^{ijk}H_{ijk}}
\label{krl-5}
\end{equation}
 Proceeding just
 as in \cite{bc}, the polarization matrix $\varepsilon = \{\varepsilon^{ij} \}$ can be brought to the following maximally reduced form
\begin{equation}
\varepsilon = \{\varepsilon^{ij} \} = \left(
\begin{array}{ccccc}
0 & 0 & 0 & 0 & 0 \\
0 & 0 & \varepsilon^{12} & \varepsilon^{13} & 0 \\
0 & -\varepsilon^{12} & 0 & \varepsilon^{23} & 0 \\
0 & -\varepsilon^{13} & -\varepsilon^{23} & 0 & 0 \\
0 & 0 & 0 & 0 & 0
\end{array}\right)
\label{ex-5}
\end{equation}
Again deleting the last row and column, one gets the polarization 
tensor$(\bar{\varepsilon})$ (\ref{bfl-2})  of the massive KR model. This is equivalent to applying
the projection operator as ${\cal P}{\varepsilon}{\cal P}$. Thus the polarization vector and  tensor of the Proca and  massive KR models,  respectively, have been
reproduced. 

Now coming to the gauge transformation properties of polarization vector (\ref{ex-2}) and polarization tensor (\ref{ex-5}) under the translational subgroup 
$T(3)$, let $W_5 (p, q, r)$ act on these objects one by one. First, acting on
$\varepsilon^{i}(\ref{ex-2})$, one gets
\begin{equation}
\varepsilon^{i} \rightarrow {\varepsilon^{\prime i}} = 
{{W_5 (p, q, r)}^{i}}_{j} \varepsilon^{j} =\varepsilon^{i} +
\delta\varepsilon^{i} = \varepsilon^{i} + ( pa_1 + qa_2 + ra_3)
\frac{p^i}{\omega}
\label{ex-6}
\end{equation}
which is indeed a gauge transformation in (4+1) dimensional Maxwell theory.
Applying the projection operator ${\cal P}$  (\ref{ex-4}) on 
(\ref{ex-6}) yields
\begin{equation}
\delta \bar{\varepsilon}^{\mu} = {\cal P}\delta\varepsilon^{i} 
= \frac{1}{\omega}( pa_1 + qa_2 + ra_3)p^{\mu} 
\label{ex-8}
\end{equation}
Here $\bar{\varepsilon}^{\mu} = (0, a_1, a_2, a_3)^T$ and, modulus the
$i$-factor, corresponds to the expression in (\ref{bfl-2}).
This is precisely how the polarization vector in
 $B \wedge F$ theory transforms under gauge transformation\cite{bc}.
In fact we can write
\begin{equation}
\delta \bar{\varepsilon}^{\mu} = {D^{\mu}}_{\nu}(p, q, r)\bar{\varepsilon}^{\nu} - \bar{\varepsilon}^{\mu}
\label{ex1-8}
\end{equation}
where, 
\begin{equation}
D(p, q, r) = \left(
\begin{array}{cccc}
1 & p & q & r \\
0 & 1 & 0 & 0 \\
0 & 0 & 1 & 0 \\
0 & 0 & 0 & 1
\end{array}
\right)
\label{ex-13}
\end{equation}
Coming next to the polarization matrix $\varepsilon = \{ \varepsilon^{ij}\}$,
its transformation law is given by,
$$\varepsilon \rightarrow \varepsilon^{\prime} = W_5(p, q, r)\varepsilon W^T_5 (p, q, r) =\varepsilon + \delta \varepsilon$$
where,
\begin{equation}
\delta \varepsilon =  \{\delta \varepsilon^{ij} \} = \left(
\begin{array}{ccccc}
0 & \alpha_1 & \alpha_2 & \alpha_3 & 0 \\
-\alpha_1 & 0 & 0 & 0 & -\alpha_1 \\
- \alpha_2 & 0 & 0 & 0 & - \alpha_2 \\
-\alpha_3 & 0 & 0 & 0 & -\alpha_3 \\
0 & \alpha_1 & \alpha_2 & \alpha_3 & 0
\end{array}
\right)
\label{ex-10}
\end{equation}
where,
$\alpha_1 = -(q\varepsilon^{12} + r\varepsilon^{13}),  ~~\alpha_2  = (p\varepsilon^{12} - r\varepsilon^{23})),  ~~\alpha_3 = (p\varepsilon^{13} + q\varepsilon^{23})$.
Again this can be easily recognized as a gauge transformation in (4+1) 
dimensional KR theory involving massless quanta, as $\delta \varepsilon^{ij}$ can be expressed as $\approx (p^{i} f^{j}(p) - p^{j}f^{i}(p))$
with suitable choice for $f^{i}(p)$, where $p^i$ is of the form (\ref{ex-3}).
 Now applying the projection operator
${\cal P}$ (\ref{ex-4}), we get
the change in the 3+1 dimensional polarization matrix
$\bar{\varepsilon} = \{ {\varepsilon}^{\mu \nu}\}$, by the formula,
$\delta \bar{\varepsilon} = {\cal P}\delta \varepsilon{\cal P}^T$. This simply
amounts to a deletion of the last row and column of $\delta\varepsilon$.
The result
can be expressed more compactly as  
\begin{equation}
\delta \bar{\varepsilon} = (D\bar{\varepsilon}D^T - \bar{\varepsilon})
\label{ex-12}
\end{equation}
where $D$ has already been defined in (\ref{ex-13}).
Again this has the precise form of gauge transformation of the
 polarization matrix 
of  $B \wedge F$ model\cite{bc}, since it can be cast in the form 
\begin{equation}
\delta \bar{\varepsilon}_{\mu \nu} = i(p_{\mu} f_{\nu}(p) - p_{\nu}f_{\mu}(p))
\label{ex-12-1}
\end{equation}
for a suitable $f_{\mu}(p)$, where the form of $p_{\mu}$ is now given by (\ref{bfl-3}).

Clearly the 
 generators $T_1 = \frac{\partial D}{\partial p}, T_2 =
\frac{\partial D}{\partial q}, T_3 = \frac{\partial D}{\partial r} $ provide a 
Lie algebra basis for the abelian group $T(3)$, as they commute
mutually. 
Note that they also satisfy
\begin{equation}
({T_1})^2 = ({T_2})^2 = ({T_3})^2 = T_1 T_2 = T_1 T_3 = T_2 T_3 = 0 
\label{ex-15}
\end{equation}
so that
\begin{equation}
 D(p, q, r) = e^{pT_1 + qT_2 + rT_3} = 1 + pT_1 + qT_2 + rT_3 
\label{ex-16}
\end{equation}
The change in the polarisation vector $\bar\varepsilon$
(\ref{ex1-8}) can be expressed as the action of a Lie algebra element,
\begin{equation}
\delta{\bar\varepsilon}^{\mu} = (pT_1 + qT_2 + rT_3){\bar\varepsilon}^{\mu}
\label{ex}
\end{equation} 
Besides this $D$ also preserves the 4-momentum of a particle at rest.
This  representation of $D(p, q, r)$ was earlier introduced in \cite{bc}.
 However, here we have shown it
 can be connected to the Wigner's little group for
massless particle in 4+1 dimensions through appropriate projection in the
intermediate steps, where the massless particles moving in  4+1 dimensions 
can be associated with a massive particle at rest in 3+1 dimensions.
Similarly the polarization vector and tensor of $B \wedge F$ theory in 
3+1 dimensions can be associated with polarization vector and polarization
 tensor of free Maxwell and KR theories in 4+1 dimensions. 

At this stage, one can ask whether the same features of dimensional
descent from higher $3+1$ dimensions will go through for the
topologically massive Maxwell-Chern-Simons theory in 2+1 dimension. 
For that let us start with free Maxwell theory in 3+1 dimension. This has 
two transverse degrees of freedom. Correspondingly the polarization vector
$\varepsilon^{\mu} $ takes the following maximally reduced 
form $\varepsilon^{\mu} = (0, a, b, 0)^T$, if the 4-momentum $p^{\mu}$ 
of the photon of energy `$\omega$' and propagating in the 3-direction
takes the following form $p^{\mu} = (\omega, 0, 0, \omega)^T$.
The generator of gauge transformation in this case is $T(2)$, which is a 
subgroup of $E(2)$. Now the representation of the 3+1 dimensional Wigner's
 little group $W_4 (p, q, \phi)$ is just obtained from $W_5(p, q, r; \psi, \phi, \eta)$ by replacing $R(\psi, \phi, \eta)$ by $R(\phi) \in SO(2)$, by 
deleting the columns and rows involving $r$ and finally setting  $r =0$ 
elsewhere. To obtain a ($4 \times 4$) representation of $T(2)$ one just sets
$\phi = 0$ in $W_4 (p, q, \phi)$.  

We are now ready to  discuss the dimensional descent from 3+1 to 2+1 dimensions.
As a first step, this
 involves reducing the 
last rows of $\varepsilon^{\mu}$ and $p^{\mu}$ to zero by using the 
projection operator ${\cal P} = diag (1, 1, 1, 0)$. 
The descended objects  $\bar{\varepsilon}^a = 
(0, a_1, a_2)^T $ ($a = 0, 1, 2$)   and $p^a = (\omega, 0, 0)^T $ 
 correspond to the polarization vector and momentum 3-vector
of a Proca quanta of mass $\omega$ at rest, of the corresponding  Proca theory
\begin{equation}
{\cal L} = -{\frac{1}{4}}{F^{ab}F_{ab}} + {\frac{\omega^2}{2}}A^{a}A_{a} 
\label{lproca}
\end{equation} 
in 2+1 dimensions. In order to discuss the gauge transformation properties 
it is essential to provide a $3 \times 3$ representation of $T(2)$ (denoted by
$\bar{D}(p, q)$) which amounts  to deleting the last row and column of $D(p, q, r)$ in (\ref{ex-13}),
\begin{equation}
\bar{D}(p, q) = \left(
\begin{array}{ccc}
1 & p & q   \\
0 & 1 & 0   \\
0 & 0 & 1   \\
\end{array}
\right)
\label{lproca-1}
\end{equation}
The corresponding generators are given by,
\begin{equation}
\bar{T}_1 = \frac{\partial \bar{D}}{\partial p} = \left(
\begin{array}{ccc}
0 & 1 & 0 \\
0 & 0 & 0 \\
0 & 0 & 0 
\end{array}
\right); \hskip 1.0cm \bar{T}_2 = \frac{\partial \bar{D}}{\partial q} =
\left( \begin{array}{ccc}
0 & 0 & 1 \\
0 & 0 & 0 \\
0 & 0 & 0
\end{array}
\right)
\label{lproca-2}
\end{equation}

Just as the Proca theory in 3+1 dimensions maps to the $B \wedge F $ theory, 
where gauge transformations  were discussed, the Proca theory in 2+1 dimensions
is actually a doublet of  
Maxwell-Chern-Simons theories,\cite{d, bw}
\begin{equation}
{\cal L} = {\cal L}_+ \oplus {\cal L}_-
\label{ex-17}
\end{equation}
where 
\begin{equation}
{\cal L}_\pm = -{\frac{1}{4}}{F^{ab}F_{ab}} \pm {\frac{{\theta
}}{2}}{\epsilon}^{abc}A_{a}{\partial}_{b}A_{c}	
\label{ex-18}
\end{equation}
with each of ${\cal L}_+$ or ${\cal L}_-$ being a topologically massive gauge
 theory. The mass of the MCS quanta is $\omega = |\theta|$, where $\omega$
 is the parameter entering in (\ref{lproca}). We can therefore study the gauge transformation generated in this 
doublet.
The polarization vector for 
${\cal L}_\pm $, with only one degree of freedom for each of ${\cal L}_+$ and
${\cal L}_-$, has been found to be\cite{g, bcs}
\begin{equation}
\bar{\varepsilon}^{a}_\pm = \frac{1}{\sqrt 2} (0, 1, \pm i)^T
\label{ex-19}
\end{equation}
while the 3-momentum $p^a$ obviously takes the same form as in the Proca model.
In analogy with (\ref{ex}), here also one can write,  
\begin{equation}\delta \bar{\varepsilon}^{a} = {(p\bar{T}_1 + q\bar{T}_2)^a}_b \bar{\varepsilon}^{b}
\label{ex-21}
\end{equation}
where $\bar{\varepsilon}^{a} = (0, a_1, a_2 )^T$
 is the polarization vector for the Proca theory in 2+1 dimensions.
Had the Proca theory been a gauge theory, (\ref{ex-21}) would
have represented a gauge transformation, as it can be written as 
\begin{equation}
\delta \bar{\varepsilon}^{a} = \frac{pa_1 + qa_2}{\omega}p^a
\label{ex-23}
\end{equation}
But since Proca theory is not a gauge theory, we can only study the gauge transformation properties of each of the doublet ${\cal L}_\pm$ (\ref{ex-18})
individually. First note that the 
 Proca polarization vector $\bar{\varepsilon}^{a}$ is just a linear 
combination of the two real orthonormal canonical vectors $\varepsilon_1$
and $ \varepsilon_2$ where, 
\begin{equation}
\bar{\varepsilon}^{a} = a_1 \varepsilon_1 + a_2 \varepsilon_2; \hskip 1.0 cm 
\varepsilon_1 = (0,1,0)^T,
\varepsilon_2 = (0,0,1)^T
\label{ex-24}
\end{equation}
Correspondingly the generators $\bar{T}_1$ and $\bar{T}_2$  (\ref{lproca-2}), 
form an 
orthonormal basis as they satisfy  $tr(\bar{T}_a^{\dagger}\bar{T}_b) = \delta_{ab}$. Furthermore, 
\begin{equation}
\bar{T}_1\varepsilon_1 = \bar{T}_2\varepsilon_2 = (1, 0, 0)^T
= \frac{p^a}{\omega}, \hskip 1.0 cm \bar{T}_1\varepsilon_2 = \bar{T}_2\varepsilon_1 = 0 
\label{ex-24-1}
\end{equation}
On the other hand, the  polarization vectors
$\bar{\varepsilon}^{a}_+$ and $\bar{\varepsilon}^{a}_-$ (\ref{ex-19})
also provide an orthonormal basis(complex) in the plane as
\begin{equation}
(\bar{\varepsilon}^{a}_+)^{\dagger}(\bar{\varepsilon}^{a}_-) = 0; 
\hskip 1.0cm (\bar{\varepsilon}^{a}_+)^{\dagger}(\bar{\varepsilon}^{a}_+) = 
(\bar{\varepsilon}^{a}_-)^{\dagger}(\bar{\varepsilon}^{a}_-) = 1
\label{ex-25}
\end{equation}
and can be obtained from the above mentioned canonical ones by appropriate  
$SU(2)$ transformation. This suggests that we consider the following
 orthonormal basis for the Lie algebra of $T(2)$:
\begin{equation}
\bar{T}_\pm = \frac{1}{\sqrt{2}}(\bar{T}_1 \mp i\bar{T}_2) = \frac{1}{\sqrt{2}}\left( \begin{array}{ccc}
0 & 1 & \mp i \\
0 & 0 & 0 \\
0 & 0 & 0 \end{array} \right)
\label{ex-26}
\end{equation}
instead of $\bar{T}_1$ and $\bar{T}_2$. Note that they also satisfy relations similar to
the (1-2) basis, 
\begin{equation}
tr(\bar{T}_+^{\dagger}\bar{T}_+) = tr(\bar{T}_-^{\dagger}\bar{T}_-) = 1; tr(\bar{T}_+^{\dagger}\bar{T}_-) = 0 
\label{ex-27}
\end{equation}
One can now easily see that 
\begin{equation}
\bar{T}_+{\varepsilon}_+ = \bar{T}_-\varepsilon_- = \frac{p^a}{\omega}, \hskip 1.0cm \bar{T}_+{\varepsilon}_- = \bar{T}_-\varepsilon_+ = 0
\label{ex-27-1}
\end{equation}
analogous to (\ref{ex-24-1}). Furthermore,
\begin{equation}
\delta \bar{\varepsilon}^{a}_\pm = \alpha_{\pm}\bar{T}_{\pm} \bar{\varepsilon}^{a}_\pm = 
 \frac{\alpha_{\pm} }{\omega}p^a 
\label{ex-28}
\end{equation}
This indicates that $\bar{T}_\pm$ - the generators of the Lie algebra of $T(2)$ in the 
rotated (complex)basis -  
generate independent  gauge transformations in ${\cal L}_{\pm}$ respectively. One
therefore can understand how the appropriate representation
 of the generator of gauge 
transformation in the doublet of MCS theory can be obtained from higher
3+1 dimensional Wigner's group through dimensional descent. 
A finite gauge transformation is obtained by integrating (\ref{ex-28}) i.e., exponentiating the corresponding Lie algebra element. This gives two
representations of Wigner's little group for massless particles in $2+1$
dimensions, which is isomorphic to ${\cal R}\times {\cal Z}_2$, although here we
are just considering the component which is connected to the identity,
\begin{equation}
G_\pm (\alpha_\pm) = e^{\alpha_\pm \bar{T}_\pm} = 1 + \alpha_\pm \bar{T}_\pm = \left(
\begin{array}{ccc}
1 & \frac{\alpha_\pm}{\sqrt{2}} & \mp i \frac{\alpha_\pm}{\sqrt{2}} \\
0 & 1 & 0 \\
0 & 0 & 1 
\end{array}
\right)
\label{ex-29}
\end{equation}
Note that $G_\pm (\alpha_\pm) $ generates gauge transformation in the 
	doublet ${\cal L}_{\pm}$, and are related by complex conjugation. This
complex conjugation is also a symmetry of the doublet.
Such representations were earlier considered in \cite{bcs}.

Another aspect of this doublet structure is illuminated by observing that,
in the 3+1 dimensional case, the photon has two helicity states $(\pm 1)$,
as follows from topological considerations \cite{sw}.
By our dimensional descent
this corresponds to the two spin states of the Proca quanta in
2+1 dimensions. Each of these states is shared amongst the two components of 
the doublet(\ref{ex-17}). This is consistent with the fact that an explicit 
computation of the spin in the MCS theory$({\cal L}_\pm)$ (\ref{ex-17})
 leads to the same values\cite{djt}.

We conclude by observing the unified picture that emerges from this analysis. 
It might be recalled that Wigner's little group for massive and massless 
particles can be regarded as having a common origin. This is because the $E(2)$
like little group  for massless particle is obtained from the $SO(3)$ like 
little group for massive particles in the infinite momentum/zero mass limit\cite{hks}.
It is essentially a consequence of the group contraction method connecting
these two groups\cite{iw}. In this paper we showed the common origin of the gauge 
generators for usual(i.e., massless)  or topologically massive gauge theories. 
The representations in the latter are obtained by a dimensional descent 
from one higher dimension of the representation in the former. Following the 
same ideas it was possible to interpret the spin quantum number of massive 
particles as the helicity quantum number of massless particles in one 
higher dimension. As a bonus
the structure of the Proca model(in 2+1 dimensions) as a doublet of 
Maxwell-Chern-Simons theories gets manifested. This is distinctive from the 
Proca model in 3+1 dimensions which gets mapped to the $B\wedge F$ model and is devoid of
any doublet structure.  Finally, the fact that the gauge generators
annihilated the physical states was very similar to the Dirac Hamiltonian
formulation\cite{pd} where the Gauss operators, which act as gauge generators,
 share
the same property. It would be worthwhile to pursue this connection between
the Wigner's little group and the Gauss operators.

\end{document}